\documentclass[
    ,final            
  ]
  {aipproc}
\usepackage{txfonts}
\usepackage{graphicx}
\usepackage{natbib}
\usepackage{aas_macros}
\bibpunct{(}{)}{;}{a}{}{,}

\layoutstyle{6x9}

\begin{document}

\title{Methane T-Dwarf Candidates in the Star Forming Region IC~348}


\author{A. Burgess}{
  address={Laboratoire d'Astrophysique,
Observatoire de Grenoble, Universit\'e J. Fourier, CNRS, BP 53, 38041
  Grenoble Cedex 9, France}
}

\author{J. Bouvier}{
  address={Laboratoire d'Astrophysique,
Observatoire de Grenoble, Universit\'e J. Fourier, CNRS, BP 53, 38041
  Grenoble Cedex 9, France}
}

\author{E. Moraux}{
  address={Laboratoire d'Astrophysique,
Observatoire de Grenoble, Universit\'e J. Fourier, CNRS, BP 53, 38041
  Grenoble Cedex 9, France}
}

\begin{abstract}
IC~348 is a young (t$\sim$3Myr) and nearby (d$\sim$340pc) star forming
region in the Perseus molecular cloud. We performed a deep imaging
survey using the MEGACAM (z-band) and WIRCAM (JHK and narrowband
CH${_4}$ on/off) wide-field cameras on the Canada-France-Hawaii
Telescope. From the analysis of the narrowband CH${_4}$ on/off deep
images, we report 4 T-dwarf candidates, of which 3 clearly lie within
the limits of the IC~348 cluster. An upper limit on the extinction was
estimated for each candidate from colour-magnitude diagrams, and found
consistent with extinction maps of the cloud. Initial comparisons with
T-dwarf spectral models suggest these candidates have a spectral type
between T3 and T5, and perhaps later, potentially making these among
the lowest mass isolated objects detected in a young star forming
region so far.
\end{abstract}

\maketitle


\section{Introduction}

In the framework of the EU Marie Curie network ``Constellation : the
origin of stellar masses'', we conducted a deep imaging survey of the
star forming region IC~348 \cite{2007AJ....134..411M} with the
aim to find isolated planetary mass objects, with a mass of a
few M$_{Jup}$, and constrain the low mass end of the Initial Mass
Function. Deep, wide-field broad-band zJHK and narrow-band CH4
on/off images were obtained at CFHT. Image analysis focused on faint
source detection, and a number of photometric tests were conducted to
ensure good photometric reliability and completeness down to
K$\sim$19.5~mag. 

\section{Selection of methane candidates}

\begin{center}
\begin{figure}[t]
\graphicspath{{images/}}
\begin{tabular}[c]{ll}
\includegraphics[width=0.5\linewidth]{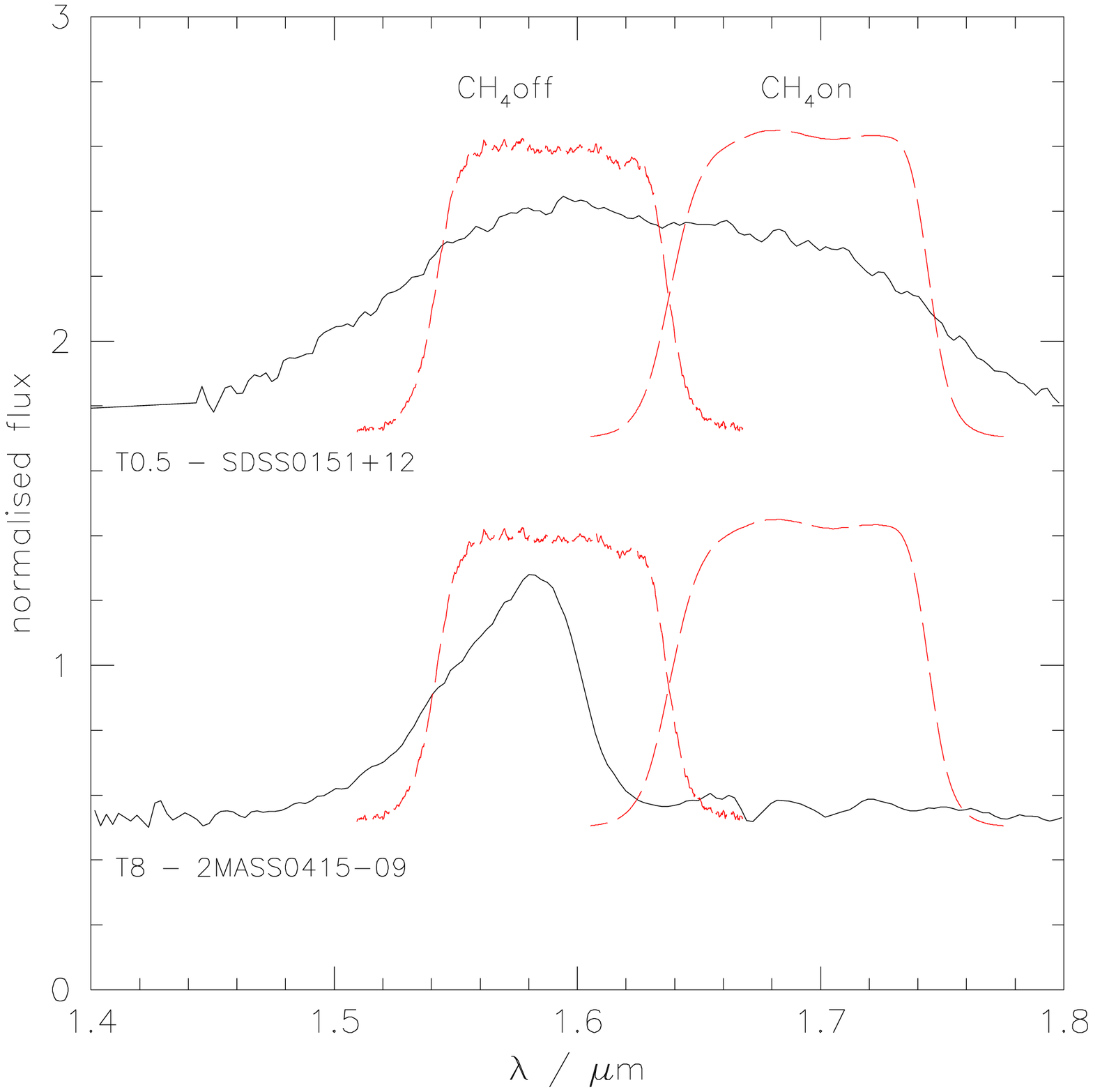} &\includegraphics[width=0.5\linewidth]{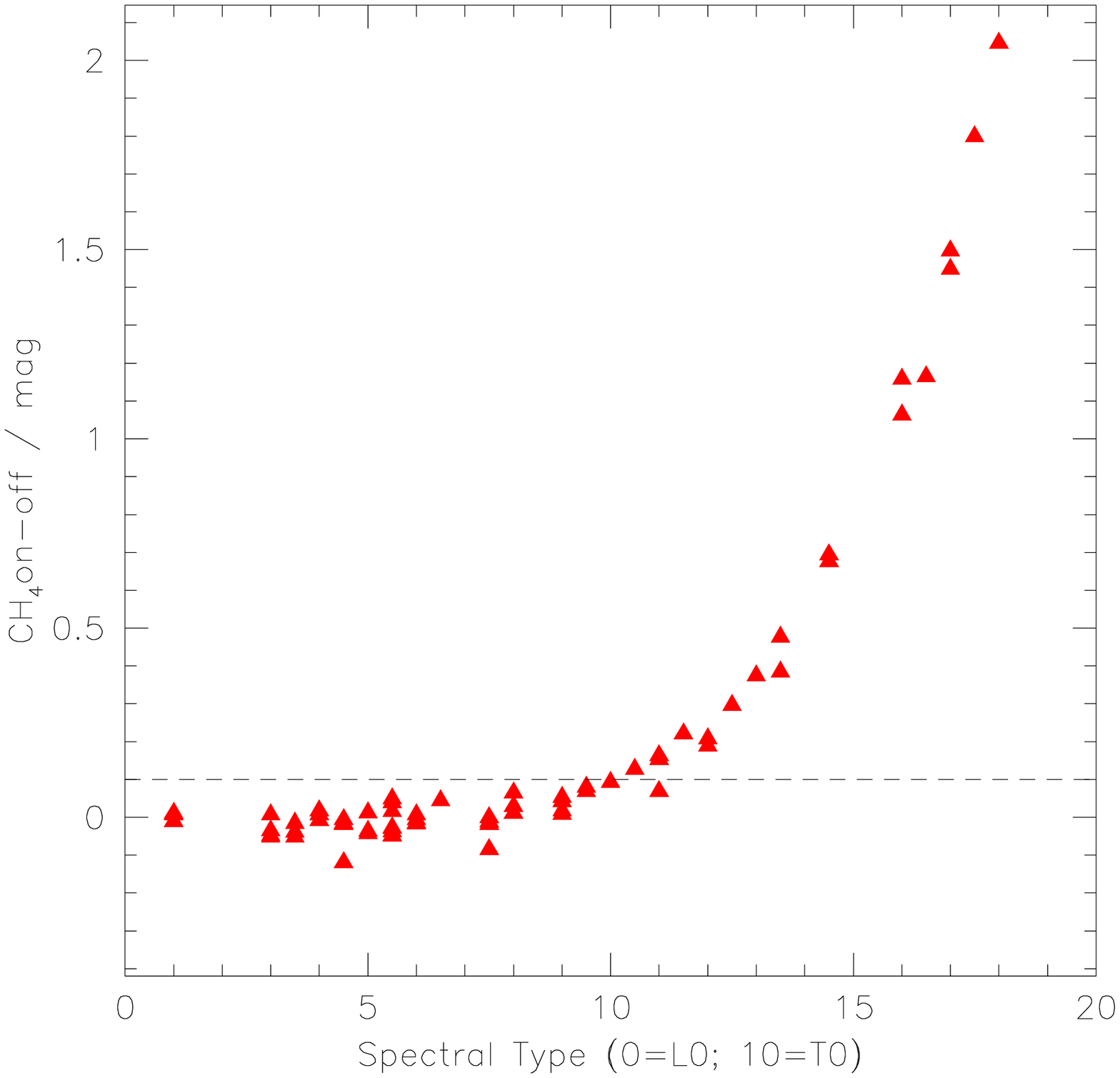} \\
\end{tabular}
\caption{{\it Left~:} Spectra of field T-dwarfs overlain by the WIRCAM
  CH$_4$ on/off filter transmission curves (dashed line). The
  CH${_4}$on (1.69$\mu$m) filter is centered on the methane absorption
  band while the CH${_4}$off (1.58$\mu$m) filter measures the nearby
  pseudo-continuum. T-dwarfs with deep methane absorption are expected
  to have larger CH${_4}$on-CH${_4}$off colours than non-methane
  dwarfs. {\it Right~:} CFHT CH${_4}$on-CH${_4}$off synthetic colours
  versus spectral type for field L1-T8 dwarfs. The L/T transition
  occurs at CH${_4}$on-CH${_4}$off$\sim$0.1~mag for field dwarfs
  (dashed line). }
\label{ch4syn}
\end{figure}
\end{center}

CH${_4}$off (1.58$\mu$m) and CH${_4}$on (1.69$\mu$m) narrowband
filters were used to detect methane absorption bands that develop in
the atmosphere of the coolest (T$\leq$1200K) objects, the
so-called T-dwarfs (cf. Fig.~\ref{ch4syn}).  We convolved observed
low resolution spectra of field L1-T8 dwarfs with MEGACAM/WIRCAM
filters in order to compute the expected CH${_4}$on-CH${_4}$off
colours as a function of spectral type. The results are shown in
Figure~\ref{ch4syn}, with the L/T dwarf transition occurring at
CH${_4}$on-CH${_4}$off$\simeq$0.1~mag for field dwarfs.

\begin{figure}[ht!]
\setlength{\unitlength}{1cm}
\centering
\graphicspath{{images/}}
\begin{tabular}[c]{ll}
\includegraphics[width=0.5\linewidth]{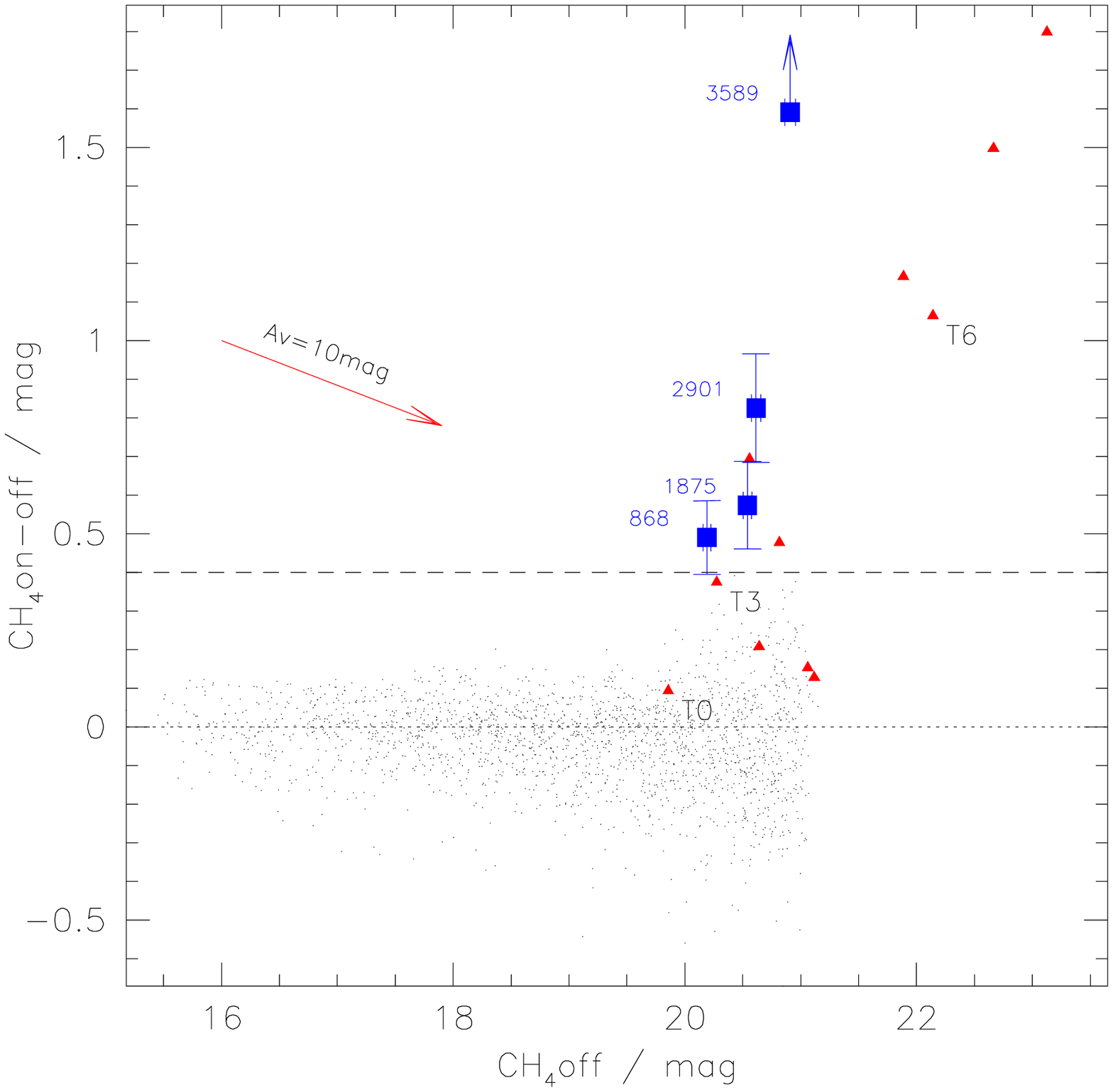} & \includegraphics[width=0.5\linewidth]{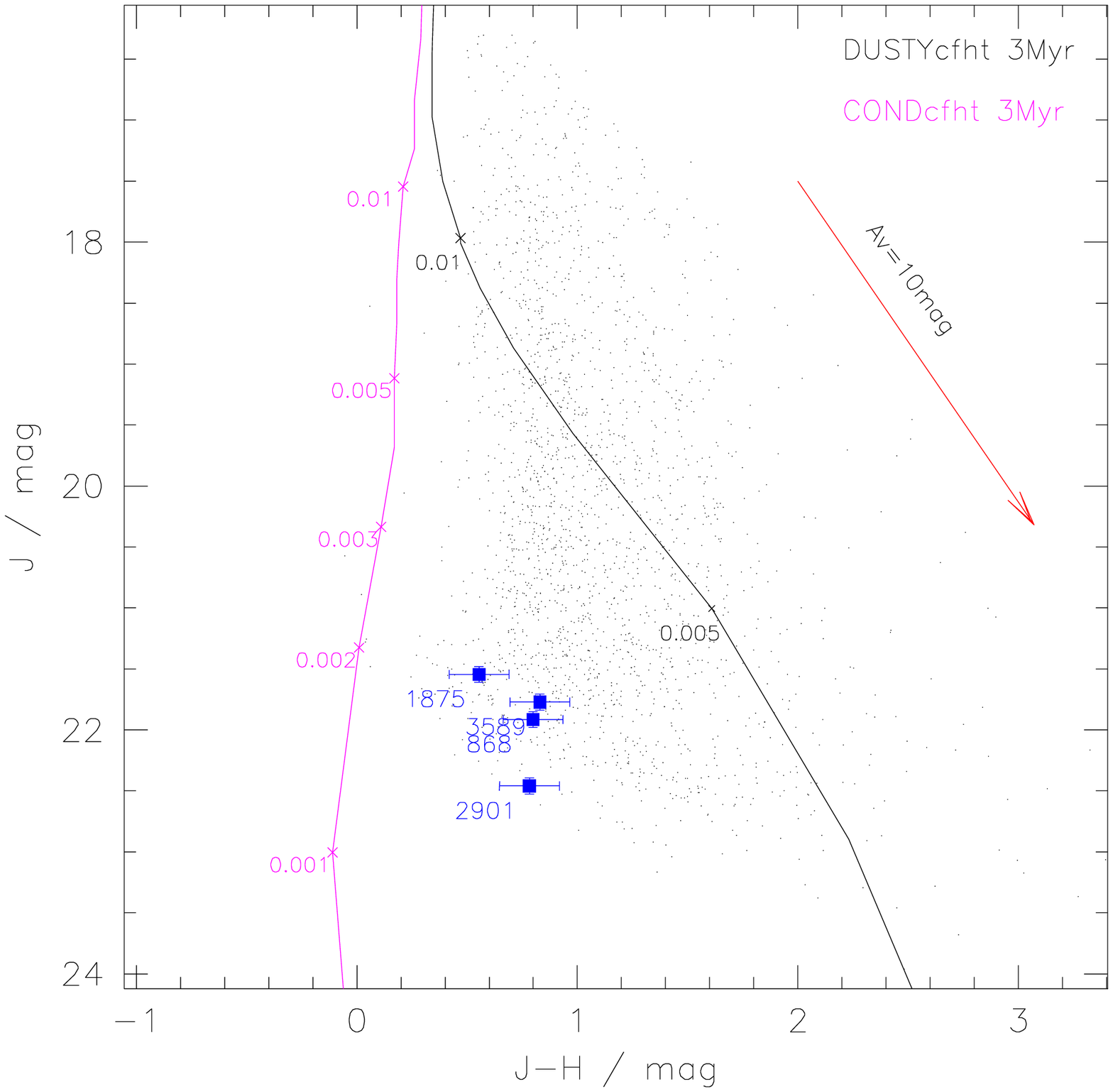} \\
\end{tabular}
\caption{{\it Left~:} Observed CH${_4}$on-CH${_4}$off colours versus
  CH${_4}$off magnitude for stars in our survey. The 4 candidate
  T-dwarfs are shown as blue squares. The field T0-T7 dwarf sequence,
  shifted to a distance of 340~pc, is shown for reference (red
  triangles). At an age of 3~Myr, IC~348 T-dwarfs are expected to be
  brighter than field dwarfs, as are the 4 identified candidates. An
  extinction vector of A$_V$=10~mag is shown. Note the asymmetric
  distribution of the CH${_4}$on-CH${_4}$off colours in this diagram,
  with more field objects with negative values, due to
  extinction. {\it Right~:} J versus (J-H) colour-magnitude
  diagram. Dusty (black) and Cond (magenta) 3~Myr isochrones are shown
  as solid lines labelled with mass (M$_\odot$). An A$_V$=10~mag
  extinction vector is shown. IC~348 T-dwarf candidates are shown as
  blue squares. They lie between Dusty and Cond isochrones as expected
  for intermediate T-dwarfs.  }
\label{onoff}
\end{figure}

Stellar-like objects were detected on the deep CH${_4}$off image and
PSF photometry was applied to both the CH${_4}$on and CH${_4}$off
images. In order to conservatively account for the photometric error,
we selected T-dwarf candidates as having
CH${_4}$on--CH${_4}$off~$\geq$~0.4~mag, which corresponds to a
spectral type of about T3 or later (cf. Fig.~\ref{ch4syn}). With this
selection limit, 135 T-dwarf candidates were initially selected. Only
4 remained, however, after visual inspection of the candidates on the
images, the other 131 being image artefacts (cross-talk, bad pixels,
etc). The 4 T-dwarf candidates are shown in a CH${_4}$on-CH${_4}$off
versus CH${_4}$off diagram (Fig.~\ref{onoff}), along with all the
other stellar-like objects detected in the images. Cand-3589 remained
undetected on the CH${_4}$on image and has been plotted using the
CH${_4}$on detection limit ($\simeq$22.5~mag). The 4 candidates stand
out in this diagram at 4$\sigma$ or more above the rms photometric
error (0.12 mag), as measured by the dispersion of the background population in
the same magnitude bin.

\section{Properties of candidate T-dwarfs}

An estimate of the extinction towards each candidate is required to
derive their absolute magnitude and, using model isochrones, mass. An
upper limit on the extinction was obtained from the J versus (J-H)
diagram shown in Figure~\ref{onoff} by projecting the candidates back
onto the COND isochrone. The results are summarised in Table
\ref{table:5} and concur with extinction maps of IC 348 by
\cite{1993BaltA...2..214C} and \cite{2003AJ....125.2029M} which
indicate 4$\leq$A$_V$$\leq$20~mag for cloud members. Table
\ref{table:5} also lists the candidates magnitude and their spectral
type estimated from Figure~\ref{onoff}. Note, however, that the
spectral classification is uncertain, as the comparison between IC~348
methane dwarfs and much older field T-dwarfs in Fig.~\ref{onoff}
assumes that their widely different gravity does not impact on their
methane colours.


\section{Cloud Membership}

The probability of one of the candidates being a foreground field
T-dwarf projected against the cloud instead of being a bona fide
IC~348 member can be estimated from the expected number density of
T3-5.5 dwarfs in the solar neighourhood, $\sim$1 per 740~pc$^{-3}$
\citep{2008ApJ...676.1281M}. The footprint of the CH${_4}$ image is
$\sim$0.11~deg$^2$, which at the distance of IC~348 ($\sim$340pc)
equates to less than one expected foreground T3-T5.5 dwarf in the
corresponding volume. Background field T-dwarfs cannot contaminate our
sample either as, at a distance of 340~pc or more, they would be at
least 1 mag fainter than our candidates in the K-band, even neglecting
extinction. Furthermore, 3 of our 4 candidates lie within 7$^{\prime}$
of the cluster centre (cf. Table~1), the half-mass radius of which is
5$^{\prime}$. We conclude that at least 3 of the 4 proposed T-dwarf
candidates are probable IC 348 members, with a mass of a few jupiter
masses according to theoretical models at an age of $\sim$3~Myr.

\begin{table}
\caption{Photometry, estimated extinction (upper limit) and spectral
type for the 4 T-dwarf candidates. The distance of each candidate from
the cluster centre
(03$^{\mathrm{h}}$44$^{\mathrm{m}}$34$^{\mathrm{s}}$;
+32$^{\circ}$09$^{\prime}$48$^{\prime\prime}$, J2000) is listed in
the last column. When the candidate remained undetected in one filter,
the magnitude listed is the detection limit, indicated with a '$*$'.
}  
\label{table:5}      
\centering                          
\begin{tabular}{l l l l l l l l l l}        
\hline                 
Object & z & J & H & K  & CH${_4}$off & CH${_4}$on &A${_V}$/mag & Sp.T. & Dist.\\
&&&&&&-CH${_4}$off\\
\hline 
868 & 23.83 & 21.92 & 21.12 & 19.9 & 20.19 & 0.49 & 6.4 & T3 & 13.4$^{\prime}$ \\
1875 & 23.30 & 21.55 & 20.99 & 20.14 & 20.54 & 0.57& 4.2 & T3.5 & 7.0$^{\prime}$ \\
2901 & $\geq$26.2$^*$ & 22.46 & 21.68 & 20.10 & 20.61 & 0.82& 6.6 & T4 & 4.1$^{\prime}$ \\
3589 & 25.22 & 21.77 & 20.94 & 20.26 & 20.91 & $\geq$1.59$^*$ & 6.6 & $\geq$T6 & 1.5$^{\prime}$ \\
\hline                                   
\end{tabular}
\end{table}

\section{Conclusion}

We report the detection of 4 methane dwarf candidates in the young
star forming region IC~348. We tentatively estimate a spectral type
in the range T3-T5 or even later, which would make these candidates 
amongst the least massive isolated objects ever detected in a star
forming region. Although follow up observations are clearly needed,
spectroscopic confirmation will be difficult owing to the faintness of
the candidates.

\begin{theacknowledgments}
This research is supported by the Marie Curie Research Training
Network "CONSTELLATION" under grant
no. MRTN-CT-2006-035890\footnote{www.constellationrtn.eu/wiki/index.php/Main\_Page}. This
research has made use of the NASA/ IPAC Infrared Science Archive,
which is operated by the Jet Propulsion Laboratory, California
Institute of Technology, under contract with the National Aeronautics
and Space Administration. This research has also made use of the
SIMBAD database, operated at CDS, Strasbourg, France. Many thanks also
goes to P. Delorme for computing the empirical sequence of L1-T8 field
dwarfs used in the figures of this paper.
\end{theacknowledgments}





\end{document}